\documentclass{article}
\usepackage{spconf,amsmath,graphicx}

\usepackage{enumitem}
\setlist{nosep, leftmargin=14pt}

\usepackage{mwe} 

\usepackage{color} 
\usepackage{colortbl}
\usepackage{cases}
\definecolor{light-gray}{gray}{0.6}
\definecolor{lavender}{rgb}{0.5,0.5,1.0}

\usepackage{amsfonts} 
\usepackage{booktabs} 
\usepackage{multirow}

\title{\textit{S\MakeLowercase{urf}}NN: J\lowercase{oint} R\lowercase{econstruction} \lowercase{of} M\lowercase{ultiple} C\lowercase{ortical} S\lowercase{urfaces} \lowercase{from} M\lowercase{agnetic} R\lowercase{esonance} I\lowercase{mages}} 
\name{Hao Zheng, Hongming Li, Yong Fan}
\address{Department of Radiology, Perelman School of Medicine, University of Pennsylvania, Philadelphia USA}
\begin{document}
%
\maketitle
\begin{abstract}
To achieve fast, robust, and accurate reconstruction of the human cortical surfaces from 3D magnetic resonance images (MRIs), we develop a novel deep learning-based framework, referred to as \textit{Surf}NN, to reconstruct simultaneously both inner (between white matter and gray matter) and outer (pial) surfaces from MRIs. Different from existing deep learning-based cortical surface reconstruction methods that either reconstruct the cortical surfaces separately or neglect the interdependence between the inner and outer surfaces, \textit{Surf}NN reconstructs both the inner and outer cortical surfaces jointly by training a single network to predict a midthickness surface that lies at the center of the inner and outer cortical surfaces. The input of \textit{Surf}NN consists of a 3D MRI and an initialization of the midthickness surface that is represented both implicitly as a 3D distance map and explicitly as a triangular mesh with spherical topology, and its output includes both the inner and outer cortical surfaces, as well as the midthickness surface. The method has been evaluated on a large-scale MRI dataset and demonstrated competitive cortical surface reconstruction performance.
\end{abstract}
\section{Introduction}
\label{sec:intro}

\begin{figure}[t]
  \centering
  \centerline{\includegraphics[width=0.95\columnwidth]{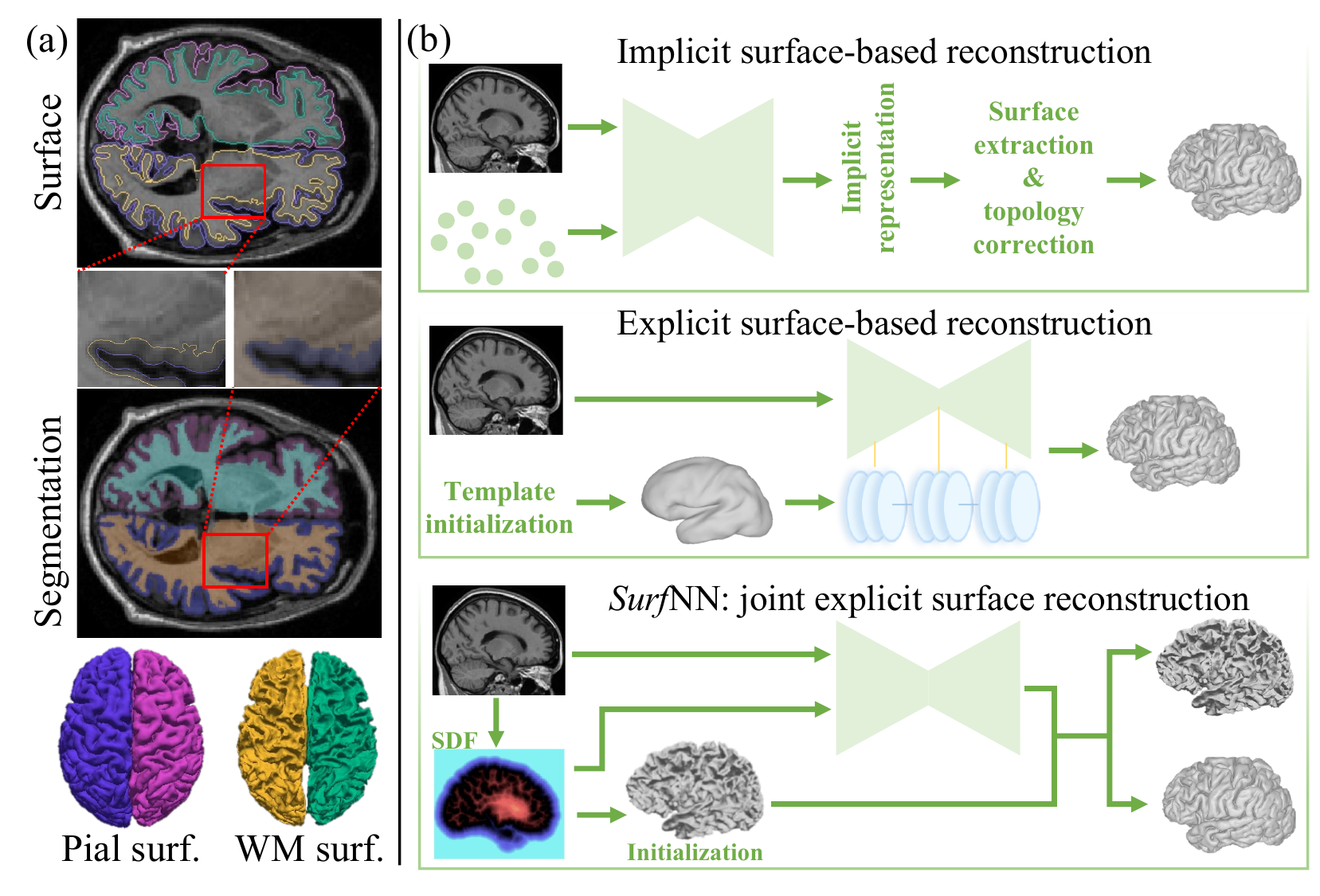}}
\vspace{-0.4cm}
\caption{(a) It is challenging to reconstruct the cortical surfaces with sub-voxel accuracy due to partial volume effect of MRI data. (b) Illustration of implicit and explicit surface-based reconstructions and comparison with our \textit{Surf}NN.}
\vspace{-0.3cm}
\label{fig:teaser}
\end{figure}

Cortical surface reconstruction plays an important role in surface-based analyses of the cerebral cortex. 
Well-established cortical surface reconstruction (CSR) methods could achieve promising performance~\cite{fischl2012freesurfer, ZengIEEE} but suffer from high computational cost and often demand manual editing to achieve sub-voxel accuracy (Fig.~\ref{fig:teaser}(a)).

\begin{figure*}[!hbt]
  \centering
  \centerline{\includegraphics[width=\linewidth]{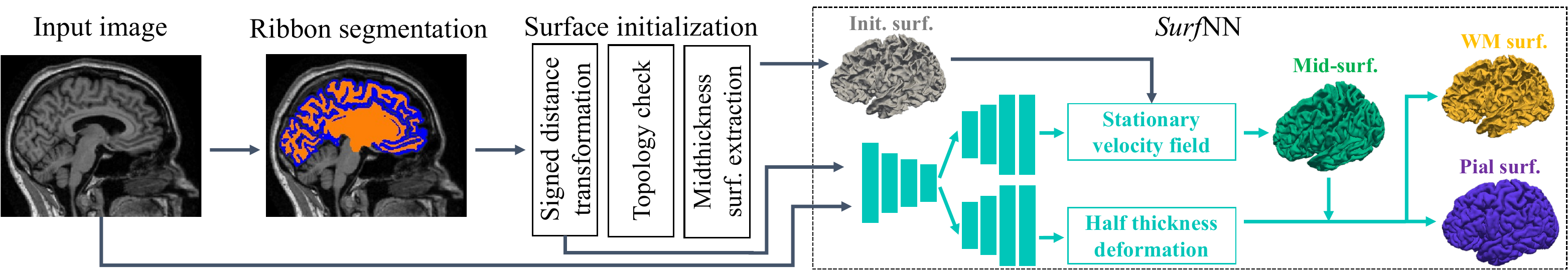}}
\vspace{-0.4cm}
\caption{An overview of our \textit{Surf}NN framework.}
\label{fig:overview}
\end{figure*}

Recent studies have demonstrated that deep learning (DL) methods could attain accurate cortical surface reconstruction while being hundreds of times faster than the conventional methods~\cite{cruz2021deepcsr, ren2022fast, lebrat2021corticalflow,santacruz2022cfpp, ma2021pialnn, bongratz2022vox2cortex, hoopes2022topofit,ma2022cortexode}. 
These DL methods utilize neural networks to predict cortical surfaces that are represented either implicitly as distance functions or explicitly as meshes (Fig.~\ref{fig:teaser}(b)).
For example, DeepCSR~\cite{cruz2021deepcsr} utilizes occupancy fields and signed distance functions (SDFs) and FastCSR~\cite{ren2022fast} employs level sets, all representing the cortical surfaces implicitly and generating surface meshes using post-processing procedures.  
By contrast, several DL methods have been developed to deform an initialization surface mesh (typically a triangular mesh) to a target mesh which can be either a white matter (WM) cortical surface (between white matter and gray matter) or a pial surface (outer cortical surface). 
Particularly, CorticalFlow methods~\cite{lebrat2021corticalflow,santacruz2022cfpp} adopt diffeomorphic deformation modules to learn a series of diffeomorphic deformations from an input MRI for deforming a template mesh towards a target cortical surface. 
CortexODE~\cite{ma2022cortexode} generates an initialization surface from an image segmentation result and then applies two models to learn diffeomorphic deformation flows for shrinking the initialization surface to a WM surface and then expanding the obtained WM surface to a pial surface subsequently. 
PialNN~\cite{ma2021pialnn} deforms an input WM surface to a pial surface by learning a one-to-one mapping between the vertices in WM and pial surfaces. 
Vox2Cortex~\cite{bongratz2022vox2cortex} uses both convolution and graph neural networks to extract hybrid features and implicitly considers the relationship between WM and pial surfaces. 
TopoFit~\cite{hoopes2022topofit} uses geometric and image features to fit a topologically-correct surface to the WM surface. 
Although these DL methods have achieved enormous success in CSR, they typically ignore the interdependence between inner and outer cortical surfaces even if both WM and pial surfaces are reconstructed jointly. Moreover, they often employ complex DL architectures, involving both convolutional neural networks (CNNs) and graph neural networks (GNNs), to learn image and surface features \emph{separately} for CSR.

We propose a novel cortical surface reconstruction method, referred to as \textit{Surf}NN, to reconstruct both WM and pial surfaces \emph{simultaneously} by learning to deform an initialization surface to halfway between the WM and pial surfaces, i.e., the midthickness surface, and at the same time to predict half cortical thickness that is used to reconstruct WM and pial surfaces from the estimated midthickness surface. 
In contrast to the existing DL methods that use both CNNs and GNNs, \textit{Surf}NN learns a diffeomorphic deformation field using CNNs from an input 3D MRI image and an initialization surface embedded in a 3D map as a signed distance function (SDF) to deform the initialization surface represented as a triangular mesh with spherical topology. The initialization surface mesh is generated from a cortical ribbon segmentation result followed by topology correction. 
\textit{Surf}NN is optimized to achieve the following goals: 
1) the initialization surface is topologically correct and as close as to the midthickness surface to retain cortical surfaces' anatomy and reduce the difficulty of deformation learning; 2) the network accurately predicts the deformations for the midthickness, WM, and pial surfaces; and 3) the deformation field is diffeomorphic to generate the surfaces with spherical topology.  
Our experiments on an ADNI dataset have demonstrated that \textit{Surf}NN can generate cortical surfaces with high accuracy and computational efficiency, superior to conventional and DL methods.

\section{Methods}
\label{sec:method}
As shown in Fig.~\ref{fig:overview}, \textit{Surf}NN framework consists of a midthickness surface initialization component, a fully convolutional network to generate a deformation field, and a half cortical thickness (HCT) module to reconstruct WM and pial surfaces simultaneously.

\subsection{Midthickness Surface Initialization}

Anatomically, the cerebral cortex has the topology of a 2D convoluted sheet with an average thickness of $\sim$$2.5mm$~\cite{fischl2012freesurfer}, thus the inner and outer cortical surfaces are coupled and separated by the cortical ribbon. Using the midthickness surface as an initialization has two advantages: 
1) The interdependence between WM and pial surfaces can be explicitly encoded via coupled surface learning; 2) Deforming a surface from the midthickness surface is less challenging than from an arbitrary surface, which may further reduce the learning difficulty and increase the accuracy. 
Our experimental results show that WM and pial surfaces can be accurately reconstructed given the midthickness surface and thickness of each vertex as long as the mesh is sufficiently dense (see Table~\ref{tab:meshden}). 

Thanks to the advancement in deep learning-based segmentation~\cite{zheng2020annotation,li2021acenet} and large-scale public neuroimaging datasets~\cite{jack2008alzheimer}, accurate segmentation of the ribbon of the cortex (i.e., the filled interior area of WM and pial surfaces) can be obtained with a DL segmentation model. 
Given an input brain MRI volume $I\in\mathbb{R}^{D\times H \times W}$, we use a 3D U-Net~\cite{ronneberger2015u} to learn a WM segmentation map $M_{w}\in\mathbb{R}^{D\times H \times W}$ and a GM segmentation map $M_{g}\in\mathbb{R}^{D\times H \times W}$ (see Fig.~\ref{fig:overview}). 


Based on a predicted segmentation map $M_{w}$, we use a distance transform algorithm to generate a signed distance function (SDF): $d(v_i)=SDF(v_i)$ is the minimal Euclidian distance of voxel $v_i\in I$ to the boundary voxels. The SDF is a volumetric level set where voxels with values equal to zero represent the surface boundaries and voxels with negative or positive values encode their distances to the surface boundaries inward or outward, respectively.
Similarly, we can generate an SDF for the pial (gray matter) surface $M_{g}$. 
By adding the WM and GM SDFs, we obtain a new level set and its 0-level defines an implicit midthickness surface. A fast topology check and correction~\cite{ma2022cortexode} is then applied to the midthickness surface to ensure its spherical topology.
Then the Marching Cubes algorithm is used to extract the initialization surface parameterized by a triangular mesh $\mathcal{S}_0$. 

\subsection{Coupled Reconstruction of Cortical Surfaces}

\begin{figure}[!hbt]
  \centering
  \centerline{\includegraphics[width=\columnwidth]{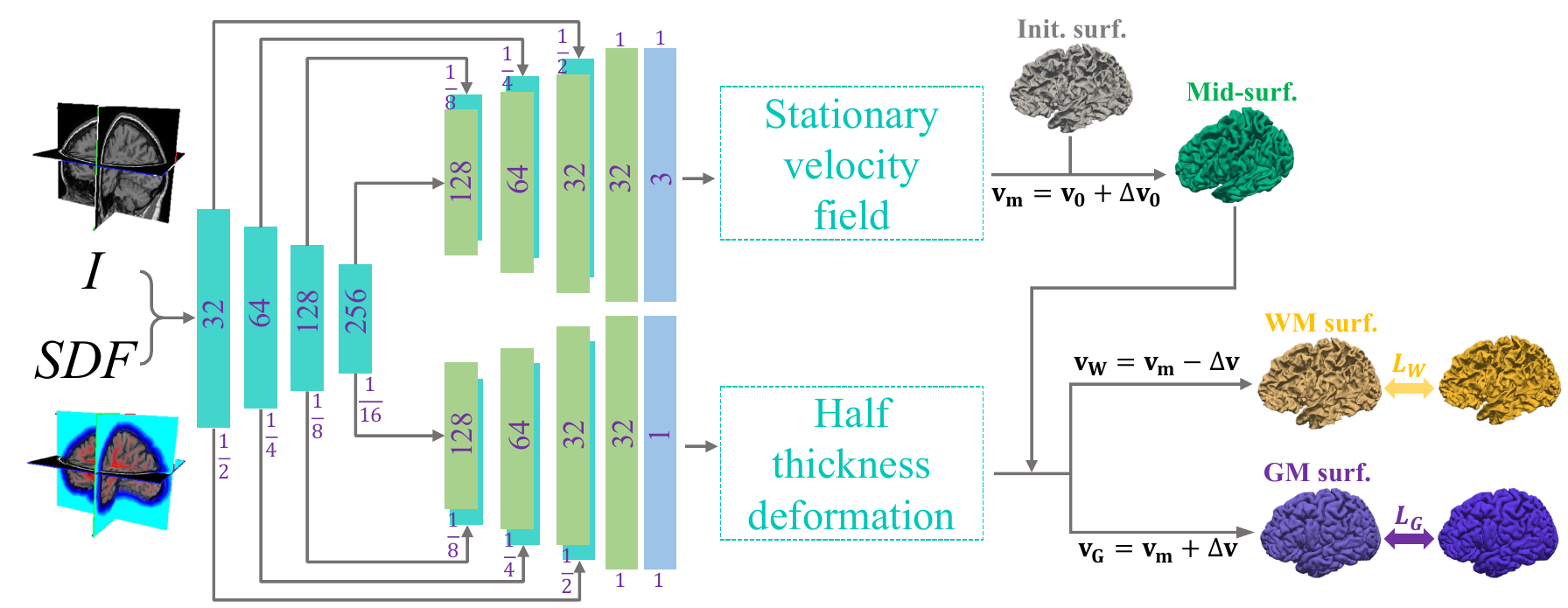}}
\vspace{-0.4cm}
\caption{Detailed structure of \textit{Surf}NN.}
\label{fig:network}
\end{figure}

Unlike the existing methods that utilize CNNs to learn image features from MRI data and GNNs to learn geometry features from surface meshes, \textit{Surf}NN learns both image and geometry features jointly from the input MRI and the SDF of initialization midthickness surface using a unified FCN to deform the initialization surface mesh. 
The unified FCN facilitates effective knowledge distillation mutually from both the 3D MRI volume and the SDF-based implicit surface. 
In order to reconstruct WM and pial surfaces robustly, \textit{Surf}NN learns to deform the initialization surface to the midthickness surface by optimizing its distances to WM and pial surfaces to be the same (i.e., half of the cortical thickness). 
Thus, we design a Y-shape network with two branches to learn an SVF-based deformation field and the half cortical thickness (HCT) jointly, as illustrated in Fig.~\ref{fig:network}.

\noindent
{\bf \textit{Surf}NN.} 
Given an initialization surface $\mathcal{S}_0\subset \mathbb{R}^3$, a neural network is optimized to learn a diffeomorphic deformation field that maps coordinates of $\mathcal{S}_0$ to that of the target midthickness surface, $\mathcal{S}_{Mid}$. 
%
The diffeomorphic field can be parameterized with the integral of a stationary velocity field (SVF), $\phi=\Phi(\mathbf{v})$, such that the surface topology is preserved and the deformation invertibility can be maintained:  
\begin{equation} 
\frac{\partial \phi_t}{\partial t} = \mathbf{v}(\phi_t),
\end{equation}
where $\mathbf{v}$ is a SVF and $\phi_t$ is the deformation field at step $t$. 
The integration of SVF, $\Phi(\mathbf{v})$, can be computed by using scaling and squaring method \cite{li2022mdreg} to obtain the deformation field. 

Once the midthickness surface is available, both WM and pial surfaces can be estimated by deforming it inward or outward by half of the cortical thickness. 
Specifically, for each vertex $\mathbf{p}\in\mathcal{S}_{Mid}$, an HCT value, $\Delta p$, is estimated such that $\mathbf{p}\prime=\mathbf{p}-\Delta p\cdot \mathbf{n} \in \mathcal{S}_W$ (WM surface) and $\mathbf{p}\prime\prime=\mathbf{p}+\Delta p\cdot \mathbf{n} \in \mathcal{S}_G$ (pial surface), where $\mathbf{n}$ the unit normal. The unit normal at each vertex is estimated based on its neighboring faces on the mesh. The HCT value, $\Delta p$, as a value related to the cortex thickness, is estimated with the lower branch in Fig.~\ref{fig:network} and sampled according to vertex coordinates of $\mathcal{S}_{Mid}$. In such a way, WM and pial surfaces are explicitly coupled in the network training. 

In summary, \textit{Surf}NN has an encoder and two decoders. The encoder consists of strided convolutions with different numbers of filters to reduce the spatial dimensions in half at each layer and to extract contextual information from both an input 3D MRI image and an SDF of the initialization surface. Each decoder consists of four deconvolutional layers concatenated with features learned in the encoding stages via skip connections, and one convolution layer to operate on the finest spatial scales and predict the SVF and HCT, respectively.


\noindent
{\bf Loss functions.} 
\textit{Surf}NN is optimized to minimize distances of the vertices between the predicted surface meshes $\mathcal{S}_W$ (and $\mathcal{S}_G$) and their corresponding ground truth (GT) meshes $\mathcal{S}_*$ by the bidirectional Chamfer distance~\cite{lebrat2021corticalflow}: 
\begin{equation} 
\small
\mathcal{L}_{chW} = \sum_{\mathbf{p}\in\mathcal{S}_W} \min_{\mathbf{p}_*\in\mathcal{S}_{W_*}}\lVert \mathbf{p}-\mathbf{p}_* \rVert_2^2 + \sum_{\mathbf{p}_*\in\mathcal{S}_{W_*}} \min_{\mathbf{p}\in\mathcal{S}_W}\lVert \mathbf{p}_* - \mathbf{p} \rVert_2^2 ,
\end{equation} 
where $\mathbf{p}$ and $\mathbf{p}_*$ are the coordinates of vertices on meshes. We can compute $\mathcal{L}_{chG}$ analogously. 

We also add edge length and normal consistency regularization terms to facilitate robust learning of the surfaces: $\mathcal{L}_{el} = \sum_{\mathbf{p} \in V}\sum_{\mathbf{q} \in \mathcal{N}(\mathbf{p})} \lVert \mathbf{p}-\mathbf{q} \rVert_2^2$, $\mathcal{L}_{nc} = \sum_{e \in E, f_0 \cap f_1 = e }(1 - cos(\mathbf{n}_{f_0}, \mathbf{n}_{f_1}))$, 
where $\mathbf{v}$ is a vertex on the predicted mesh (i.e., $\mathcal{S}=(V, E)$), $\mathcal{N}(\mathbf{v})$ is the set of all neighboring vertices of $\mathbf{v}$, $e$ is an edge on the predicted mesh, $f_0$ and $f_1$ are $e$'s two neighboring faces along with their unit normals $\mathbf{n}_{f_0}$ and $\mathbf{n}_{f_1}$. 

In summary, we combine all the surface geometric losses to jointly optimize the model: $\mathcal{L} = \lambda_1 \mathcal{L}_{ch} + \lambda_2 \mathcal{L}_{el} + \lambda_3 \mathcal{L}_{nc} $, where $\lambda_i (i=1,2,3)$ are the weights to
balance loss terms. For simplicity, we let  $\lambda_1 = \lambda_2 = \lambda_3 = 1$.

\section{Experiments}
\label{sec:exp}

We evaluated \textit{Surf}NN on a large-scale MRI dataset and compared it with representative CSR methods chosen from each category as discussed in Sect.~\ref{sec:intro}, with the cortical surfaces represented as distance functions or meshes, including DeepCSR~\cite{cruz2021deepcsr}, PialNN~\cite{ma2021pialnn}, and cortexODE~\cite{ma2022cortexode}.

\begin{table*}[!hbt]
\centering
\caption{ Cortical Surface Reconstruction Accuracy on ADNI dataset. All values are in $mm$. The best ones are in bold. } 
\label{tab:main}
\small
{
\begin{tabular}{ c |c c c | c c c| c c c| c c c}
\toprule
 & \multicolumn{3}{c|}{L-Pial Surface } & \multicolumn{3}{c|}{L-WM Surface}  & \multicolumn{3}{c|}{R-Pial Surface}&
\multicolumn{3}{c}{R-WM Surface} 
\\   
Method   & CD  & AD & HD  & CD  & AD & HD   & CD  & AD & HD   & CD  & AD & HD    \\  \cline{1-13} 
DeepCSR~\cite{cruz2021deepcsr}  & 0.986 & 0.684 & 1.387 & 0.977 & 0.672 & 1.219 & 1.013 & 0.693 & 1.412 & 0.996 & 0.680 & 1.226     \\
PialNN~\cite{ma2021pialnn} & 0.675 & 0.516 & 1.113 & $\backslash$ & $\backslash$ & $\backslash$ & 0.653 & 0.491 & 1.044 & $\backslash$ & $\backslash$ & $\backslash$ \\  
cortexODE~\cite{ma2022cortexode} & 0.452 & {\bf 0.203} & {\bf 0.434} & 0.436 & 0.189 &  0.414 & 0.461 & {\bf 0.199} & {\bf 0.423} & 0.441 & 0.191 & 0.418 \\ 
\textit{Surf}NN$_{w/o SDF}$ & 0.471 & 0.256 & 0.571  & {0.349} & {0.177} & {0.381} & {0.446} & 0.259 & 0.582 & {0.345} & {0.175} & {0.380}  \\ 
\textit{Surf}NN  & {\bf 0.407} & {0.231} & {0.523} & {\bf 0.290} & {\bf 0.143} & {\bf 0.306} & {\bf 0.408} & {0.233} & {0.525} & {\bf 0.289} & {\bf 0.141} & {\bf 0.304} \\  
\bottomrule

\end{tabular}
}
\end{table*}

\noindent
{\bf Dataset.}
We evaluated our method on a large-scale dataset, {\bf ADNI}~\cite{jack2008alzheimer}, consisting of 747, 70, and 113 subjects for training, validation, and testing, respectively. 
The MRI volumes were aligned rigidly to the MNI152 template and clipped to the size of $192 \times 224 \times 192$ at $1mm^3$ isotropic resolution. 
The surrogate ground truth (GT) surfaces were generated by FreeSurfer v7.2.0~\cite{fischl2012freesurfer} for training and evaluation. The MRI scans' intensity values were normalized to $[0,1]$ and the coordinates of the vertices were normalized to $[-1,1]$. 

\noindent
{\bf Implementation details.}
Pytorch was used to develop all DL models and an NVIDIA P100 GPU with 16 GB memory was used for all experiments. 
The 3D U-Net~\cite{ronneberger2015u} for segmentation of ribbons was trained for 200 epochs using Adam optimization with a learning rate $10^{-4}$ and achieved an average Dice index of 0.96 on the testing set. 
The \textit{Surf}NN was trained for 400 epochs to reconstruct WM and pial surfaces of each hemisphere. 
The surface meshes had ${\sim}130k$ vertices. 

\noindent
{\bf Evaluation metrics.}
We utilized three distance-based metrics to measure the surface reconstruction error: 
Chamfer distance (CD)~\cite{lebrat2021corticalflow}, 
average absolute distance (AD)~\cite{cruz2021deepcsr}, and 90th percentile Hausdorff distance (HD)~\cite{cruz2021deepcsr}. They were computed bidirectionally in millimeters ($mm$) and over point clouds of ${\sim}130k$ points uniformly sampled from the predicted and target surfaces. A lower distance means a better result. 

\begin{figure}[htb]
  \centering
  \includegraphics[width=\columnwidth]{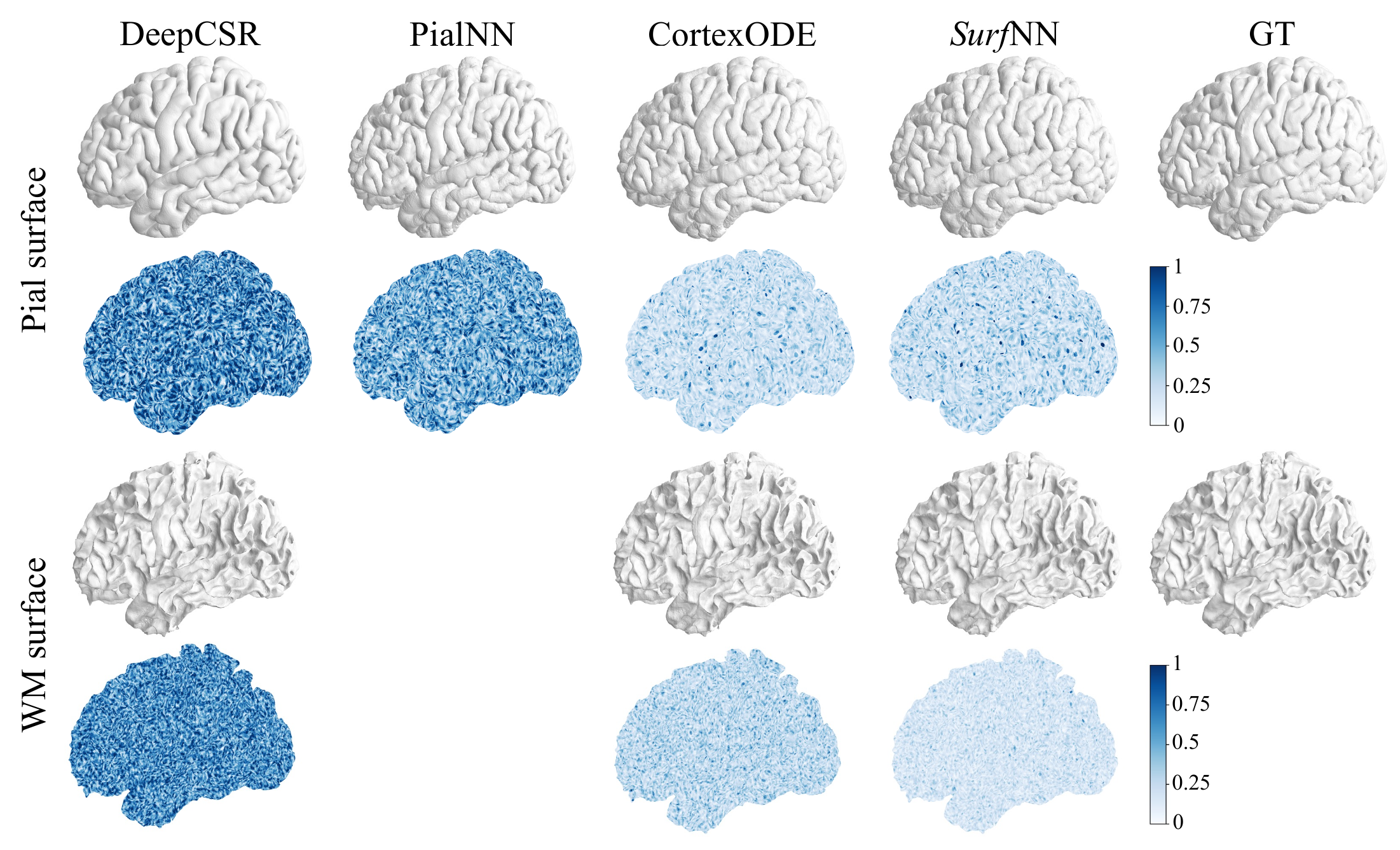}
\vspace{-0.4cm}
\caption{ 1st \& 3rd rows: reconstructed cortical surfaces. 2nd \& 4th rows: The average absolute distance (AD, in $mm$) map between the predicted surfaces and the ground truth.}
\label{fig:res}
\vspace{-0.3cm}
\end{figure}

\noindent
{\bf Main experimental results.}
The experimental results are summarized in Table \ref{tab:main}. 
\underline{First}, all DL methods for modeling surface meshes explicitly outperformed DeepCSR by a large margin. 
\underline{Second}, PialNN generated pial surfaces with an average AD value much higher than those generated by cortexODE, and our \textit{Surf}NN obtained $55.2\%$ improvement on L-pial surface (i.e., $0.231mm$ \emph{v.s} $0.516mm$). 
\underline{Third}, our \textit{Surf}NN obtained comparable or superior performance to cortexODE. For example, the mean average AD of L-Pial and L-WM surfaces of \textit{Surf}NN was $0.008mm$ lower than that of cortexODE (i.e., $0.187mm$ \emph{v.s} $0.195mm$). 
Besides, \textit{Surf}NN has two extra advantages: 1) \emph{one} model for reconstructing both WM and pial surfaces simultaneously; and 2) jointly generating the midthickness surface and estimating the cortex thickness that can be utilized in statistical analyses of brain atrophy. 
Representative results are visualized in Fig.~\ref{fig:res}.

\noindent
{\bf Runtime.} 
It took $4min$, $3.2s$, and $1.8s$ for DeepCSR~\cite{cruz2021deepcsr}, cortexODE~\cite{ma2022cortexode}, and \textit{Surf}NN to reconstruct both WM and pial surfaces, respectively, given an initialized surface. 

\noindent
{\bf Impact of the SDF.} 
To evaluate the effectiveness of SDF in the \textit{Surf}NN, we conducted an ablation study by removing the SDF from the network input, denoted as \textit{Surf}NN$_{w/oSDF}$. As shown in Table~\ref{tab:main}, SDF played an important role in facilitating the feature extraction of \textit{Surf}NN, such as boosting the L-WM surface AD by $19.2\%$ (i.e., $0.143mm$ \emph{v.s} $0.177mm$).

\noindent
{\bf Impact of input surface density.}
We also evaluated how the surface reconstruction accuracy changes with the surface mesh density. As shown in Table.~\ref{tab:meshden}, the performance of \textit{Surf}NN improved with the increasing density of the surface meshes and can achieve promising accuracy when surface meshes had more than $130k$ vertices.

\begin{table} 
\centering
\caption{ Surface reconstruction w/ different mesh densities. }
\label{tab:meshden}
\small
{
\begin{tabular}{ c|ccc|ccc } 
\toprule

\multirow{ 2}{*}{ \shortstack{Num. of \\ vertices} }  & \multicolumn{3}{c|}{L-Pial Surface } & \multicolumn{3}{c}{L-WM Surface} 
\\   
  & CD & AD & HD & CD & AD & HD  \\ \cline{1-7}
$40k$    & 0.569 & 0.344 & 0.725 & 0.471 & 0.243 & 0.589 \\ 
$130k$   & 0.407 & 0.231 & 0.523 & 0.290 & 0.143 & 0.306  \\

 
\bottomrule

\end{tabular}
}
\end{table}

\section{Conclusion}
\label{sec:conclusion}

We developed a novel DL method, \textit{Surf}NN, to reconstruct both WM and pial surfaces simultaneously from MRIs in a coupled way from an initialization midthickness surface, facilitating accurate reconstruction of the cortical surfaces. Moreover, \textit{Surf}NN also generates the midthickness surface and an estimation of cortical thickness, which can be used in statistical analyses of brain atrophy.

\section{Compliance with ethical standards}
\label{sec:ethics}

This study was conducted retrospectively using public data~\cite{jack2008alzheimer}. 

\section{Acknowledgments}
\label{sec:ack}

This work was supported in part by the NIH grant AG066650 and EB022573.



\bibliographystyle{IEEEbib}
\bibliography{refs}

\end{document}